\documentclass[3p,preprint,times]{elsarticle}

\usepackage{graphicx}
\usepackage{multirow}
\usepackage{ecrc}
\usepackage{graphicx}
\usepackage{algpseudocode} 
\usepackage{multicol}
\usepackage{comment}

\volume{00}

\firstpage{1}

\journalname{Computers \& Security}

\runauth{X Yuan et al.}

\jid{COSE}

\jnltitlelogo{Procedia Computer Science}

\CopyrightLine{2023}{Published by Elsevier Ltd.}

\usepackage{amssymb}
\usepackage[ruled,vlined]{algorithm2e}

\usepackage[figuresright]{rotating}
\usepackage{amsmath}

\usepackage{color}
\newcount\Comments  
\definecolor{darkgreen}{rgb}{0,0.5,0}
\definecolor{purple}{rgb}{1,0,1}
\newcommand{\kibitz}[2]{\ifnum\Comments=1\textcolor{#1}{#2}\fi}

\begin{document}

\begin{frontmatter}



\dochead{\kibitz{red}{\textbf{Accepted by Computers \& Security}}}

\title{A Simple Framework to Enhance the Adversarial Robustness of Deep Learning-based Intrusion Detection System}


\author[1]{Xinwei Yuan}
\ead{symor@seu.edu.cn}

\author[1]{Shu Han}
\ead{220205123@seu.edu.cn}

\author[4]{Wei Huang\corref{cor1}}
\ead{huangwei@pmlabs.com.cn}

\author[1]{Hongliang Ye}
\ead{yehongliang@seu.edu.cn}

\author[4]{Xianglong Kong}
\ead{kongxianglong@pmlabs.com.cn}

\author[5]{Fan Zhang}
\ead{17034203@qq.com}

\cortext[cor1]{Corresponding author}

\address[1]{Southeast University, Nanjing, China}
\address[4]{Purple Mountain Laboratories, Nanjing, China}
\address[5]{National Digital Switching System And Engineering Technological Research Center, Zhengzhou, China}

\begin{abstract}
Deep learning based intrusion detection systems (DL-based IDS) have emerged as one of the best choices for providing security solutions against various network intrusion attacks. However, due to the emergence and development of adversarial deep learning technologies, it becomes challenging for the adoption of DL models into IDS. In this paper, we propose a novel IDS architecture that can enhance the robustness of IDS against adversarial attacks by combining conventional machine learning (ML) models and Deep Learning models. The proposed DLL-IDS consists of three components: DL-based IDS, adversarial example (AE) detector, and ML-based IDS. We first develop a novel AE detector based on the local intrinsic dimensionality (LID). Then, we exploit the low attack transferability between DL models and ML models to find a robust ML model that can assist us in determining the maliciousness of AEs. If the input traffic is detected as an AE, the ML-based IDS will predict the maliciousness of input traffic, otherwise the DL-based IDS will work for the prediction. The fusion mechanism can leverage the high prediction accuracy of DL models and low attack transferability between DL models and ML models to improve the robustness of the whole system. In our experiments, we observe a significant improvement in the prediction performance of the IDS when subjected to adversarial attack, achieving high accuracy with low resource consumption.
\end{abstract}

\begin{keyword}
intrusion detection system \sep adversarial example \sep adversarial detection \sep adversarial defense\sep deep learning \sep machine learning \sep classification algorithm



\end{keyword}
\end{frontmatter}

\section{INTRODUCTION}

In the past decades, the development of computer networks has made remarkable achievements. With the explosion of the network, the number of devices in the network increases exponentially, accompanied by more serious security threats to network devices. These security threats can endanger the confidentiality, integrity, and availability of assets. Therefore, it is crucial to protect networks and information systems from potential network attacks. IDS \kibitz{black}{has} emerged as one of the best choices for providing security solutions against various network intrusion attacks. IDS works effectively through two main methods: signature-based detection and anomaly-based detection \cite{b2}. Feature-based detection \cite{b3} involves analyzing and summarizing known attacks, and encoding the extracted attack features into rules or signatures. When network traffic matches the predefined rules, an alarm will be triggered. Although it can effectively prevent known threats, defense effectiveness will be greatly reduced when encountering unknown threats. In addition, anomaly-based detection \cite{b4} can mitigate the impact of unknown threats by analyzing and modeling normal traffic. It raises an alarm when encountering any network traffic that is abnormal or deviates from the baseline. 

With the rapid progress of hardware technology and the emergence of various GPUs with high computational power, many researchers are exploring the possibility of applying machine learning (ML) techniques to the detection of IDS. The common methods of ML used in IDS are decision tree (DT) \cite{b16}, Support Vector Machine (SVM) \cite{b17}\cite{b18}, and K-Nearest neighbor (KNN) \cite{b19}, but the effectiveness of ML depends largely on the complexity of the dataset and the size of the data. It would be difficult and expensive to adapt a large network architecture and complex network conditions with an ML-based IDS. Compared with the ML-based IDS, DL-based IDS has more hidden layers to obtain deeper features of the network, so it comes into the view of researchers as a new development direction of IDS. Besides, many excellent DL models have excellent performance in this field, and they account for more than 80\% of the frequency of use in research, becoming the mainstream direction of IDS development today. 

Although DL models can be used to construct high-performance IDS, the highly non-linearity of neural networks makes them sensitive to small perturbations. Consequently, attackers can manipulate the detection results of IDS through adversarial attacks. Specifically, adversarial attacks involve introducing crafted perturbations to clean examples, resulting in the generation of adversarial examples (AEs). As a result, deep neural networks(DNN) can correctly classify the clean examples but mis-classify the AEs. The adversarial attack was first proposed by Szegedy et al.\cite{b20}
In recent years, more and more studies have started to focus on adversarial \kibitz{black}{attacks} in the scenario of IDS. Wang et al.\cite{b4} first attempted to conduct a feature-level attack (FLA) on the \kibitz{black}{multilayer perceptron (MLP)} model using a white-box \kibitz{black}{(WB)} approach, and they controlled the perturbation to be bounded by $l_p$ norm ball, which is similar to settings in computer vision (CV) attack. Lin \cite{b21} pioneered the gray-box \kibitz{black}{(GB)} attack using \kibitz{black}{a generative adversarial network (GAN)} and used a generator with limited prior knowledge to generate perturbations with adversarial properties, which were superimposed on the feature dimension of the original sample to mislead the deep neural network successfully. Alhajjar \cite{b22} added a genetic algorithm (GA) to generate AEs based on Lin et al.'s scheme. Their results show that DL-based IDS can severely damage its availability when subjected to adversarial attacks. 

It is essential to establish a precise definition beforehand. Adversarial examples compromise the predictive ability of DL-based IDS, without posing an inherent threat to computer systems. Conversely, conventional attack traffic manifests its malicious nature by causing damage to computer systems, generally unable to evade detection by DL-based IDS\footnote{In this paper, we use "adversarial" to describe the perturbed input, which are misclassified by DL models, and "malicious" to describe the input traffic, which cause damage to computer systems.}. Obviously, adversarial attacks are great threats to the network system. Therefore, it is crucial to deploy defense plans for DL-based IDS\cite{b24}. There are three existing defense approaches for adversarial attacks: parameter protection, robustness optimization, and AE detection.

Parameter protection is not the optimal way to enhance the robustness of IDS against adversarial attacks since it can be bypassed in some time\cite{b25}. Additionally, robustness optimization is not practical for feature-level attacks. Therefore, we tend to train an efficient AEs detector to protect IDS from security threats. However, with the deepening of research, it is inevitable to encounter a difficult problem: how to deal with the traffic detected as an AE? As for CV, the existence of AEs can serve as evidence that the model is under attack. In such cases, it is advisable to interrupt the current process and discard the corresponding examples. However, for IDS that adopted bypass deployment, it is necessary to confirm whether these AEs will pose security threats to the computer systems protected by the IDS, and decide whether to generate an alarm. In general, AEs are generated by malicious traffic to deceive IDS to attack computer systems without triggering alarms. Similarly, they can also be generated from benign traffic, and are used to deceive IDS to generate massive false alarms, so that IDS completely loses its credibility. Therefore, it is not sufficient to judge the malicious intent of traffic solely based on the detection of AEs. We believe that an IDS should assess traffic examples from both adversarial and malicious intent perspectives.

In this paper, we attempt to address the above problem by proposing a new framework called the DLL-IDS system, which consists of three main components: DL-base IDS, LID-based AE detector, and ML-based IDS. Moreover, it provides a unique malicious intent and adversarial label for each traffic example. By observing that AEs exhibit different distribution characteristics in high-dimensional space compared to clean examples, we innovatively introduce a method based on the local intrinsic dimensionality\cite{b23}. This method utilizes the activation values of hidden layers in deep neural networks to describe the spatial features of samples. The computed value from the LID can be approximated as the dimensional space where the sample resides. Generally, AEs are expected to exist in a higher-dimensional space compared to clean examples. We can leverage the differences between the two to train a high-accuracy AE detector. Since LID characterizes the collective properties of samples, different attack methods generating AEs will exhibit similar spatial attributes. This property enables the AE detector to have a certain degree of transferability in scenarios with limited prior knowledge. After performing AE detection, further discrimination of the malicious nature of the traffic is required. We have observed that traditional ML models exhibit high robustness when faced with AEs. As a result, we have identified a ML algorithm, Label-spreading (LS), which demonstrates lower attack transferability. This is a semi-supervised learning algorithm where new samples passed to the model are treated as unlabeled samples. The labels of these samples are determined collectively by the surrounding samples, and the weight factors are determined based on the distances between them. The perturbations added during the generation of AEs cause changes in the distances between these samples and other clean samples. However, the strict constraints on the magnitude of perturbations confine the generated AEs within the manifold. The normalized Laplacian matrix further controls the influence of neighboring samples. Therefore, in this scenario, the LS model exhibits robustness against AEs. We utilize this model to make the final predictions on the maliciousness of the AEs.

The contributions of our paper are summarized as follows:
\begin{itemize}
    \item We are the first to address the specific requirements of IDS in the face of adversarial attacks and propose a comprehensive evaluation of input traffic from both malicious intent and adversarial perspectives. 
\end{itemize}
\begin{itemize}
    \item We propose a new IDS architecture against adversarial attacks. The innovation lies in introducing the concept of local intrinsic dimensionality into IDS AE detection. Our detection approach achieves relatively high accuracy under various attacks, and the detection accuracy increases as the intensity of adversarial attacks rises.
\end{itemize}
\begin{itemize}
    \item We observe varying levels of attack transferability between DL models and traditional ML models and propose a method to enhance the robustness of IDS systems against adversarial attacks by combining these two types of models. To be specific, we employ the LS machine learning method and explored the reasons behind its robustness on AEs for the first time.
\end{itemize}
\begin{itemize}
    \item Our experiments show that the accuracy of IDS is compromised to varying degrees when subjected to different \kibitz{black}{intensities} of adversarial attacks. In the worst case, the accuracy drops to only 17.9\%. In contrast, DLL-IDS maintains an accuracy of over 70\% even under similar attack scenarios, with accuracy reaching as high as 90\% in some cases. Additionally, DLL-IDS maintains a high accuracy of over 90\% on clean examples, showing no significant difference in detection capability compared to the original IDS.
\end{itemize}


\section{BACKGROUND \& RELATED WORK}
\subsection{Adversarial attacks on IDS}

With the widespread application of deep learning technology, \kibitz{black}{DL-based IDS are more flexible and efficient than traditional IDS}, which learn the extracted and constructed features, and independently complete the entire process of building a network feature behavior library, learning rules, and establishing a machine-learning model. However, the emergence and development of adversarial machine learning technology pose new challenges to DL-based IDS. \kibitz{black}{Adversarial attacks deliberately introduce imperceptible disturbances to input examples, leading to classification errors in the classifier. The samples with these added disturbances are referred to as AEs \cite{b1}.} At present, many AE generation schemes have been proposed, \kibitz{black}{with} the goal being to generate AE that satisfies both confidence and similarity constraints and to carry out attacks in different ways on different target IDS \cite{b2}. The existing similarity constraints for IDS adversarial attacks include $l_p$ ball, generative model, and packet obfuscation, and existing research on DL-based IDS against adversarial attacks primarily focuses on feature-level attacks (FLA).  The defense approach proposed in DLL-IDS also targets this type of attack, so we will not cover much discussion based on other similarity constraints and attack methods.

FLA is an attack method that directly perturbs the characteristics of the input network traffic. Instead of destroying the output layer, it destroys the characteristics of the middle layer of the model to enhance the migration of AEs. FLA typically employ black-box \kibitz{black}{(BB)}, \kibitz{black}{GB}, and \kibitz{black}{WB} attack against target IDS, with similarity constraints set within an $l_p$ norm ball. In terms of \kibitz{black}{WB} attacks, Rigaki \cite{b3} launched \kibitz{black}{WB} attacks based on fast gradient sign method (FGSM) and Jacobian-based saliency map attack (JSMA) on IDS based on different classifiers including DT, random forest (RF), SVM, Voting and MLP. Their experiments demonstrated that using substitute models could generate effective AEs, and adversarial attacks based on JSMA were more effective in significantly reducing the accuracy of IDS compared to attacks based on FGSM. Wang et al. \cite{b4} also investigated four \kibitz{black}{WB} attack methods, including JSMA, FGSM, DeepFool, and CW \cite{b5}, to evaluate the robustness of IDS based on MLP.  They compared the destruction and applicability of these four attacks and concluded that JSMA is more attractive to attackers. They also found that certain features in different adversarial attacks contribute to the generation of AEs, which need to be protected from being exploited by adversaries. Clements et al. \cite{b6} conducted a study on the robustness of a deep learning-based IDS called Kitsune for IoT networks. They used the Mirai dataset and evaluated the system's robustness against FGSM, JSMA, CW, and ENM attacks. The results showed that all of these algorithms had a success rate of 100\% in terms of integrity attacks. However, only CW and ENM showed good effectiveness in availability attacks. It was found that adversaries could create effective adversarial examples by changing an average of 1.38 traffic features. But this may lead to the generation of unrealistic adversarial features. In order to generate realistic features, Sheatsley et al. \cite{b7} proposed an adaptive JSMA (AJSMA) method to prevent violating network constraints during the extraction of network traffic features. Then they used a histogram sketch generation (HSG) algorithm to generate adversarial sketches, ensuring that the perturbations adhere to network constraints. Even in the presence of adversarial constraints, there is a certain probability of successfully generating AE that satisfies domain constraints. The attacks mentioned above are based on \kibitz{black}{WB} attacks. However, Yang et al. \cite{b8} conducted \kibitz{black}{BB} attacks on DNN models using three different methods: training substitute models, ZOO framework, and Wasserstein GAN. These attacks demonstrated the ability to generate effective AEs even when the adversary knows nothing about the classifier. Therefore, adversarial attacks are inevitably becoming a significant threat and a challenging problem for DL-based IDS. Effectively defending against adversarial attacks has become a hot research topic and will continue to be extensively studied.

\subsection{Adversarial \kibitz{black}{Defense} and Its Applicability in IDS}

Due to the potential threat posed by adversarial attacks, adversarial defense has become an important research topic in the field of deep learning. However, the current adversarial defense approaches mostly focus on CV and have not extensively addressed their applicability in the field of IDS. Existing research directions for adversarial defense can be categorized into three types: parameter protection, robustness optimization, and adversarial detection \cite{b2}. Parameter protection is commonly used to defend against model inversion attacks \cite{b9}, and its core idea is to employ gradient hiding techniques, which can be achieved through gradient discretization, stochastic gradients, and gradient vanishing or exploding. For example, Guo et al. \cite{b10} applied image transformations such as image cropping, rescaling, and bit-depth reduction to achieve gradient discretization, making the transformed images non-differentiable and internally randomized, thus achieving better defense effectiveness. However, Athalye et al. \cite{b11} proposed a method to construct effective adversarial examples by using differentiable approximations of non-differentiable functions during neural network backpropagation, thus bypassing gradient hiding. Additionally, the unique characteristics of network traffic data make it difficult to quantify, so parameter protection-based adversarial defense is not the optimal solution for achieving robustness against adversarial attacks in DL-IDS. Robustness optimization \cite{b2} is another effective defense method against adversarial attacks, aiming to enhance the robustness of classifiers to achieve the correct classification of AEs. For example, Madry et al. \cite{b12} proposed generating AEs using projected gradient descent (PGD) and using these AEs to train the model. Tramèr et al. \cite{b13} combined multiple pre-trained models with AEs generated through adversarial attacks to train the target model. 

Adversarial training is one of the most effective defense methods in robust optimization, but it is not well-suited for IDS due to the uncertainty in obtaining network attack traffic. This would require a significant amount of resources to continuously update the model in real-time. Besides, adversarial training leads to a shift in the decision boundary of the model, resulting in a decrease in prediction accuracy for clean examples. For large-scale network systems, where thousands of traffic flows may pass through the IDS per second, even a slight decrease in the accuracy for clean examples can result in hundreds of false positives.  And this is a cost that we cannot afford. The above two methods both involve the training process of the model, while adversarial detection typically involves detecting the presence of AE before the classification. Metzen et al. \cite{b14} proposed a dynamic adversary training method to enhance the ability of detectors to handle dynamic adversaries. They detect AEs by examining intermediate feature representations, where the input of the convolutional layers is passed to a secondary classifier to determine its adversarial intent. Feinman et al.\cite{b40} combined density estimation and bayesian uncertainty to detect AEs from the perspectives of manifold density and confidence space of AEs. However, the aforementioned defense methods are mostly based on CV and have not been applied to IDS. MANDA \cite{b15} is the first adversarial defense scheme proposed for IDS. Based on the clustering characteristics and positional attributes of AEs, MANDA combines manifold detection and \kibitz{black}{DB(decision boundary)}methods. Manifold detection calculates the discrepancy of sample distributions and identifies AEs if the sample is inconsistent with the model's prediction. However, using manifold detection alone as an AE detection algorithm lacks solid theoretical support. DB introduces perturbations to input examples to detect AEs located near the decision boundary. Unfortunately, it lacks adaptability to different attacks and requires generating multiple perturbed examples $x'$ for each input example $x$, which significantly reduces the efficiency of detection. To the best of our knowledge, we are the first to emphasize the complete requirements of AE defense based on IDS and the first to propose the classification problem of AEs generated by benign traffic and malicious traffic.

\section{SYSTEM MODEL AND THREAT MODEL}

\subsection{Weakness of Deep Learning}
Firstly, deep learning models have a gap between their decision boundaries and the true classification boundaries of input samples. This gap can be exploited to generate AEs that cause the model to make incorrect classifications. The presence of AEs can be influenced by the quantity and distribution of the training data.
Second, \kibitz{black}{data} often exists in high-dimensional spaces. Perturbing each dimension of the data to generate AEs will result in a collection of small perturbations for each dimension. These small perturbations can be sufficient to change the classification result\cite{b43}. 
These two characteristics determine that deep neural networks are vulnerable to adversarial attacks. For DL-based IDS, attackers can generate AEs from benign traffic to attack the model and cause it to generate a large number of false positives. They can also generate AEs from malicious traffic to hide their malicious attacks. This can pose a serious threat to the security of network systems.
\subsection{Threat model}   
In this article, the target of the simulated attacker is to cause the IDS model to classify traffic into a wrong class, regardless of whether the traffic is benign or malicious. Under this assumption, a successful attack against the IDS model should be able to reduce the model's accuracy to below the normal range. The threat model refers to the generation conditions of the AE, and among all the attacks against deep learning models, we can classify attacks based on the attacker's prior knowledge into the following three types:

\begin{itemize}
\item WHITE-BOX ATTACK: The attacker knows everything about the victim model, including the training data, model output, architecture, and weights. The attacker uses information about the model, especially the gradient, to generate AE.
\item BLACK-BOX ATTACK: The attacker cannot access the parameters of the victim model. The attacker generates AE by observing the input and output of the victim model.
\item GRAY-BOX ATTACK: The attacker knows the training data, but does not know the architecture and weight parameters of the victim model. Therefore, the attacker chooses to construct an alternative model based on the transferability property of AEs, which performs the same task as the victim model to generate AEs. 
\end{itemize}
Among all the attack categories, \kibitz{black}{WB} attacks have the strongest destructive power and have the greatest impact on model testing results. Therefore, we consider using four \kibitz{black}{WB} attack methods to attack deep learning models.

\textbf{First}, we adopt FGSM, it is the first developed $L_{\infty}$ attack that uses DL gradients to generate an AE. The core idea of the algorithm is to quickly calculate the sign of the gradient to determine the direction of the gradient descent. That is, we need to calculate the gradient direction of $y$ with respect to $x$, and then add random noise to $x$ to generate adversarial examples in the direction of the largest gradient descent. It is expressed as follows:
\begin{align}
     x^{\prime}=x+\epsilon \operatorname{sign} \nabla_{x} J(\theta, x) .
\end{align}

\textbf{Second}, we adopt basic iterative method(BIM)\cite{b44}, Based on the theory of gradient calculation for FGSM, $x$ proposed an iterative approach to finding each dimension's perturbation. At each iteration, the existing perturbation is corrected based on the previous iteration's gradient and then clipped to the range of $(0, 1)$. It is expressed as follows:
\begin{align}
\boldsymbol{X}_{0}^{a d v}=\boldsymbol{X}, \quad \boldsymbol{X}_{N+1}^{a d v}=\operatorname{Clip}_{X, \epsilon}\left\{\boldsymbol{X}_{N}^{a d v}+\alpha \operatorname{sign}\left(\nabla_{X} J\left(\boldsymbol{X}_{N}^{a d v}, y_{t r u e}\right)\right)\right\} .
\end{align}

\textbf{Third}, we adopt Deepfool\cite{b46}, a minimal perturbation attack algorithm that finds the direction and size of the minimum perturbation by calculating the orthogonal projection distance between the sample and the decision boundary. Then, it iterates to generate adversarial examples until the classifier changes its output result. This attack generates a smaller perturbation than FGSM at the same fooling ratio. It is expressed as follows:
\begin{align}
\boldsymbol{r}_{*}\left(\boldsymbol{x}_{0}\right) & :=\arg \min \|\boldsymbol{r}\|_{2} \\
& \text { subject to } \operatorname{sign}\left(f\left(\boldsymbol{x}_{0}+\boldsymbol{r}\right)\right) \neq \operatorname{sign}\left(f\left(\boldsymbol{x}_{0}\right)\right) \\
& =-\frac{f\left(\boldsymbol{x}_{0}\right)}{\|\boldsymbol{w}\|_{2}^{2}} \boldsymbol{w}.
\end{align}

\textbf{Fourth}, we adopt CW\cite{b47}, an algorithm that follows the optimization problem of L-BFGS with few changes. the optimization problem is expressed as follows:
\begin{align}
   \min _{\delta}\|\delta\|+\operatorname{cg}\left(x^{\prime}\right), \text { such that } x^{\prime} \in[0,1]^{n} .
\end{align}

It uses w to control the perturbation of the input sample: $\delta =1/2\left(\tanh(w)+1 \right)-x$. This attack is considered as one of the state-of-the-art attacks. It is first implemented to break distillation knowledge defense and it was shown it is stronger than FGSM and BIM attacks.
\subsection{System model}

\begin{figure}[htbp]
  \centering
  \includegraphics[width=0.7\linewidth]{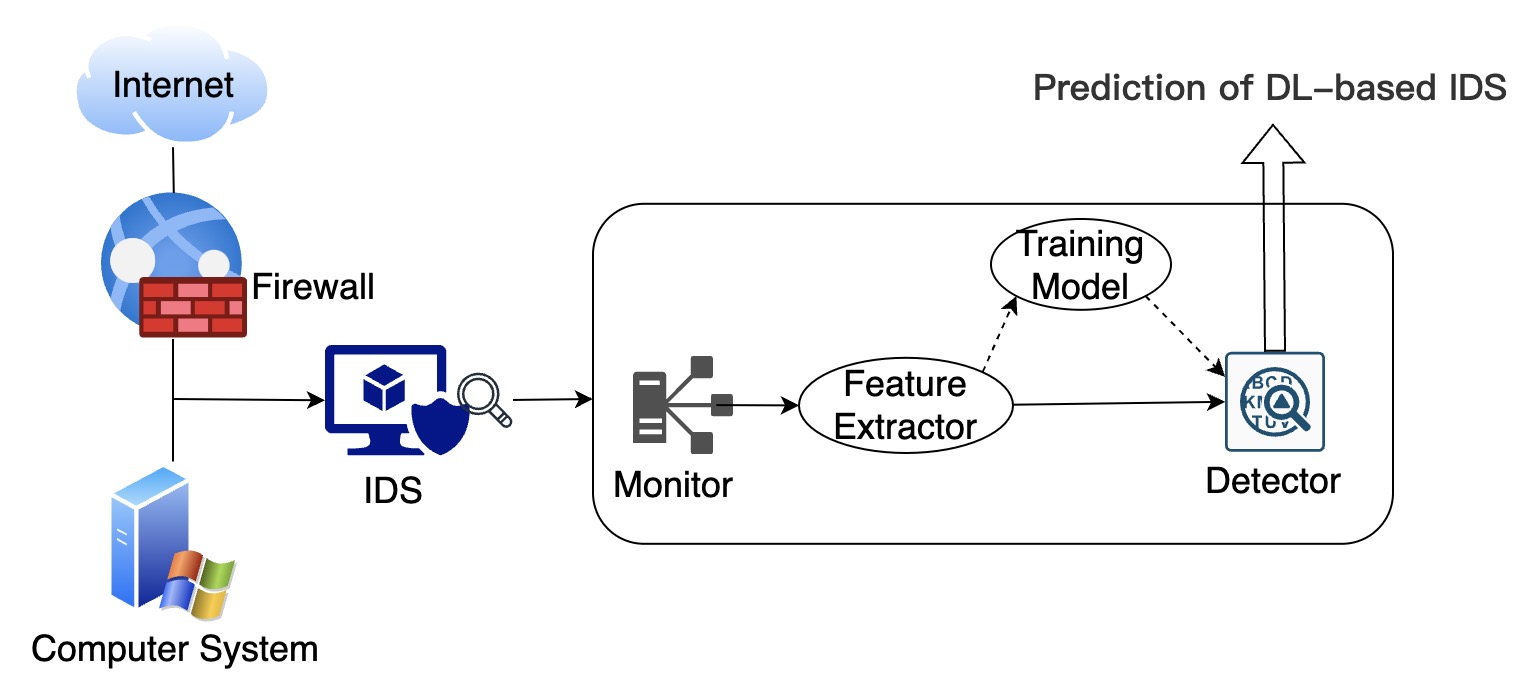} 
  \caption{System model of DL-based IDS.}
  \label{fig1}
\end{figure}
Figure \ref{fig1} shows the architecture of a typical DL-based IDS. The DL-based IDS system usually consists of four parts: monitor, feature extractor, training model, and Detector model. The monitor collects and provides information for data preprocessing and detectors. Feature extraction is the process of extracting quintuples and traffic features from the original traffic packets, and then formatting them. The training model can be considered as training a neural network model that can distinguish malicious traffic from normal traffic through a large number of original examples. The DL-based IDS model provides traffic classification and prediction through the classification results of a neural network model. In real network environments, external traffic is typically mirrored and copied after passing through a firewall and then transmitted to the IDS for analysis and detection. After the monitor receives the traffic, it is immediately transmitted to the data preprocessing module. Once the data processing is complete, the results are sent to the DL-based IDS model. It is assumed that the DL-based IDS model has already undergone pre-training. The DL-based IDS model then provides a prediction regarding the nature of the traffic, specifically whether it is malicious or not. Finally, this prediction is sent to the user. Due to the vulnerability of DL model to the adversarial attack, DL-based IDS may provide wrong prediction of adversarial traffic. We can categorize these AEs based on the maliciousness of input traffic.

The AEs of malicious traffic: to hide their true attack intent and attack payload, attackers attempt to conduct destructive attacks on the system protected by IDS by creating AEs for malicious traffic without triggering alarms. It is important to note that both the original malicious traffic and the AE generated from it are malicious in nature. Furthermore, both types of traffic can cause damage to the network system.

\begin{figure}[h]
  \centering
  \includegraphics[width=0.7\linewidth]{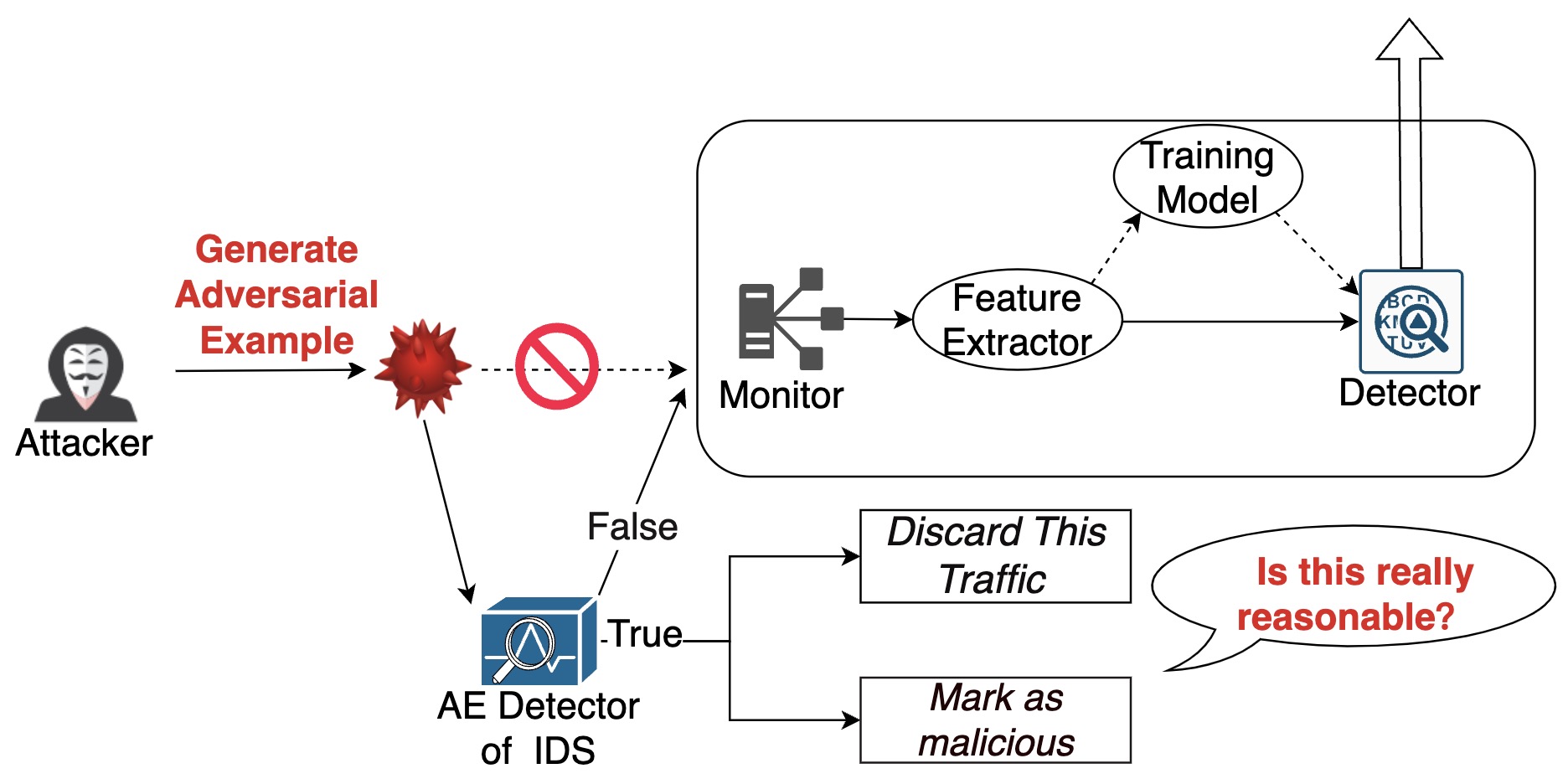}
  \caption{System model of the current AE defense for IDS}
  \label{fig2}
\end{figure}

The AEs of normal traffic: to interfere with the normal detection of IDS, attackers attempt to create AEs of clean traffic, which trigger a large number of ‘benign’ alarms in the IDS, leading to a decrease in its availability. It is worth noting that both the original benign traffic and the AEs generated from it are benign in nature. Neither type of traffic will cause any damage to the network system.

The research on adversarial defense for DL-based IDS mainly focuses on adversarial detection, and its workflow (as shown in the figure \ref{fig2}) is most similar to the approaches of AEs defense in the field of object detection. After AE detection, if adversarial attacks are detected, the traffic will be directly discarded or marked as malicious. This type of decision-making approach is acceptable in the field of object detection because the actual content in the images is far less important than the presence of adversarial attacks. Therefore, determining that the input example belongs to an AE is sufficient to consider the system under attack and terminate the current object detection process. However, in the case of network traffic being analyzed for malicious intent, it is possible for the traffic itself to possess malicious characteristics that are unrelated to adversarial attacks. We believe that IDS should predict whether any incoming network traffic is malicious. However, if the input example is an AE, it may be detected as malicious traffic. In this case, the IDS is only focusing on the subjective maliciousness of the attacker generating AEs, rather than determining whether the traffic is inherently malicious. There is a scenario: an attacker uses benign traffic to make a large number of AEs and sends these AEs into the network. For the existing IDS defense schemes\cite{b15}, these AEs will be judged as malicious traffic by IDS and trigger a large number of attack alarms. Actually, this network traffic will not cause a destructive impact on the system, and there is no risk threat to the server cluster protected behind the IDS. However, massive warning messages can seriously damage the usability and readability of IDS. And it is even necessary to manually check these alarm messages, which will inevitably cause a large waste of resources. Apparently, it is unacceptable. Based on the above problem, it is concluded that IDS should not only pay attention to whether the input examples are AEs but also focus on whether they belong to malicious traffic.

\section{DLL-IDS System}

\begin{figure}[h]
  \centering
  \includegraphics[width=0.5\linewidth]{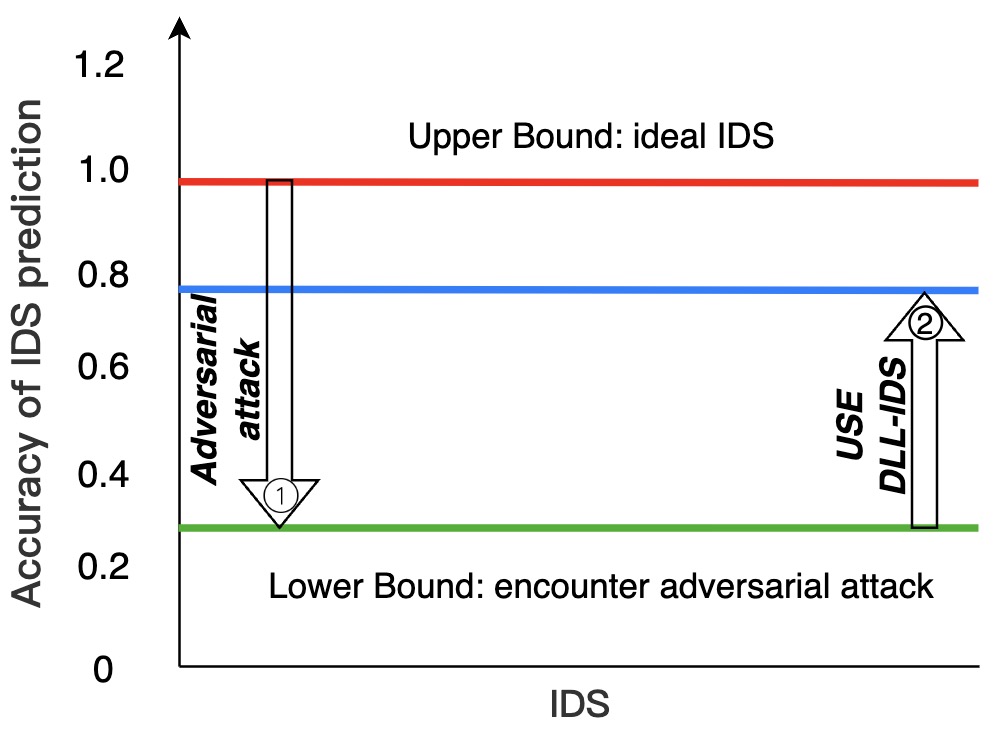}
  \caption{The system significance of DLL-IDS}
  \label{fig3}
\end{figure}

In this section, we propose a novel IDS system architecture that defends against adversarial attacks. This IDS system evaluates traffic from two perspectives: whether the input traffic is an AE and the maliciousness of the traffic itself. Figure \ref{fig3} illustrates the conceptual diagram of the proposed system, which is designed to significantly enhance the lower bound of IDS performance when under attack, leveraging the high efficiency of DL models and the great robustness of ML models. The upper bound depicted in the figure is determined by the predictive performance of the model itself on clean examples, while the lower bound is determined by the intensity of adversarial attacks. Under high-intensity attacks such as CW, the model's prediction accuracy may be less than 10\%. Our task is to design a novel IDS system architecture that can enhance the robustness of IDS against adversarial attacks. It should also meet the following requirements mentioned earlier: 1) distinguishing the maliciousness of the traffic itself, and 2) determining whether an example belongs to an AE. 

\begin{figure}[h]
  \centering
  \includegraphics[width=0.85\linewidth]{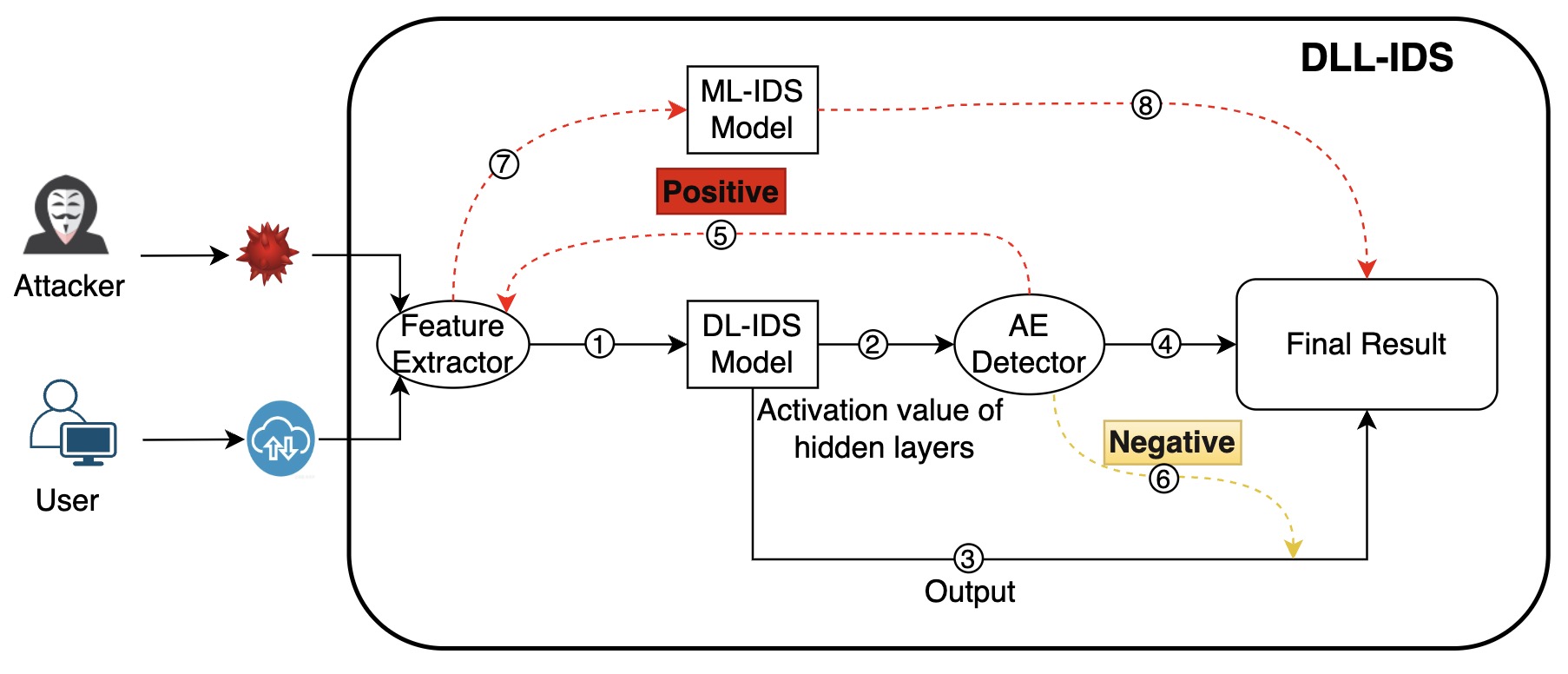}
  \caption{The realization of DLL-IDS}
  \label{fig4}
\end{figure}

To achieve this goal, we propose an IDS system architecture, as depicted in Figure \ref{fig4} and Algorithm 1. In contrast to the traditional single-model DL-based IDS, this IDS system comprises three modules: DL-based IDS, ML-based IDS, and AE detector. Whether it is a clean example or an AE, input traffic entering the DLL-IDS system will first go to the DL-based IDS. This IDS model will perform two operations: 1) the DL model predicts the maliciousness of the sample, and 2) the activation values of each layer of the sample in the DL model (used for the AE detector) are passed forward to the AE detector. The AE detector is capable of determining whether the input traffic belongs to AEs based on the activation values of the hidden layers in the previous step of the DL model and \kibitz{black}{outputting the final detection results}. The AE detector will then handle two scenarios. If the detection result is positive (input traffic is an AE), it will go back to the feature extraction step and pass the extracted features to the ML-based IDS (assuming that this ML model has a certain level of robustness against AEs). The ML-based IDS will reevaluate the maliciousness of the flow based on this model, and the prediction result of the ML-based IDS will be used as the final prediction result for the maliciousness of the traffic. If the AE detection result is negative, the prediction result of the DL-based IDS module will be trusted. Finally, \kibitz{black}{two prediction results are aggregated together and output}. Assuming that the system receives only clean examples, it is reasonable to expect that the ML-based IDS model's resources will not be utilized. Conversely, if only AEs are provided, relying on the AE detector's high detection rate can prevent interference from the DL-based IDS in the system's final decision-making process. This combination of system modules maximizes the utilization of different models and achieves optimal defense effectiveness with minimal system resource consumption. We introduce the design principles of these three modules as follows.

\begin{center}
\begin{minipage}{0.7\linewidth}
\begin{algorithm}[H]
	\caption{DLL-IDS}
	\label{alg:algorithm2}
	\KwIn{%
		\\$X$: examples of network traffic from outside \\
		$D(x)$: a pre-trained DL-based IDS model \\
        $Hid-Act(x)$: the Activation value of $x$ in each hidden layer of DL model
		$M(x)$: a pre-trained ML-based IDS model \\
		$AD(lid)$: a LID-based AE detector
	}
	\KwOut{%
		\\$isAdversarial \in \{\text{False}, \text{True}\}$ \\
		$isMalicious \in \{\text{False}, \text{True}\}$
	} 
	\BlankLine
	$Result_{adv} = []$, $Result_{mal} = []$ \\
	\ForEach{$x$ in $X$}{
		$Result_{mal} = D(x)$ \\
		$Hid-Act(x) \gets D(x)$ \\ 
		Compute $lid_{x}$ \\
		\If{$AD(lid_{x})$}{
			$Result_{adv} = \text{True}$ \\
			$Result_{mal} = M(x)$
		}\Else{
			$Result_{adv} = \text{False}$
		}
		$Result_{adv}.\text{append}(Result_{adv})$ \\
		$Result_{mal}.\text{append}(Result_{mal})$
	}
	\Return $Result_{adv}$, $Result_{mal}$
\end{algorithm}
\end{minipage}
\end{center}

\subsection{DL-based IDS}

In recent years, deep learning has been applied to IDS due to its ability to learn hierarchical network structures and extract features from multiple hidden layers\cite{b26}. DNNs serve as the fundamental architecture for deep learning, consisting of input layers, hidden layers, and output layers. Given their excellent performance in classification tasks, this study adopts DNN as a fundamental classifier for the IDS. \kibitz{black}{Within the realm of DNN, a diverse array of network architectures, including fully-connected feedforward neural network (FNN), convolutional neural network (CNN), recurrent neural network (RNN), and others, are incorporated. These architectures exhibit a complex mapping relationship between the input vector and the output vector in the network model. The input vectors represent feature vectors obtained after meticulous data preprocessing, while the output comprises a probability vector that characterizes various classification categories. The input-output relationship of the hidden layer can be represented as follows: $h = g(wx + b)$, where $g$ represents the activation function of the hidden layer, $w$ denotes the weights between the input layer and the hidden layer, and $b$ represents the biases of the hidden neurons. In this paper, the DL-based IDS has adopted both FNN and CNN models tailored to different datasets. Then, we provide an overview of the FNN model designed for the NSL-KDD dataset.} The architecture, as shown in Figure 5, the input vector is a 128-dimensional vector obtained after data preprocessing, resulting in 128 neurons in the input layer \kibitz{black}{(The reason for 128 neurons will be explained in 5.2)}. The output vector is a two-dimensional probability vector ranging from 0 to 1, representing benign or malicious traffic. The output result of the model is determined by the maximum probability value in the two-dimensional output vector. Through multiple experiments, we have found that increasing the number of layers improves the prediction accuracy for our network traffic dataset, with a relative peak achieved at five layers. The FNN model with five hidden layers, as shown in Figure 6, is chosen as the basis for the IDS in this study. \kibitz{black}{The accuracy of this FNN model for clean traffic samples of NSL-KDD is already close to 95\%. Our research focuses on the defense against adversarial attacks, instead of the factors that maximize the accuracy of the DL-based IDS}. Therefore, these DL models are selected as the DL-based IDS component in the DLL-IDS system.
\begin{figure}[htbp]
    \centering
    \begin{minipage}[b]{0.6\linewidth}
        \centering
        \includegraphics[width=1\linewidth]{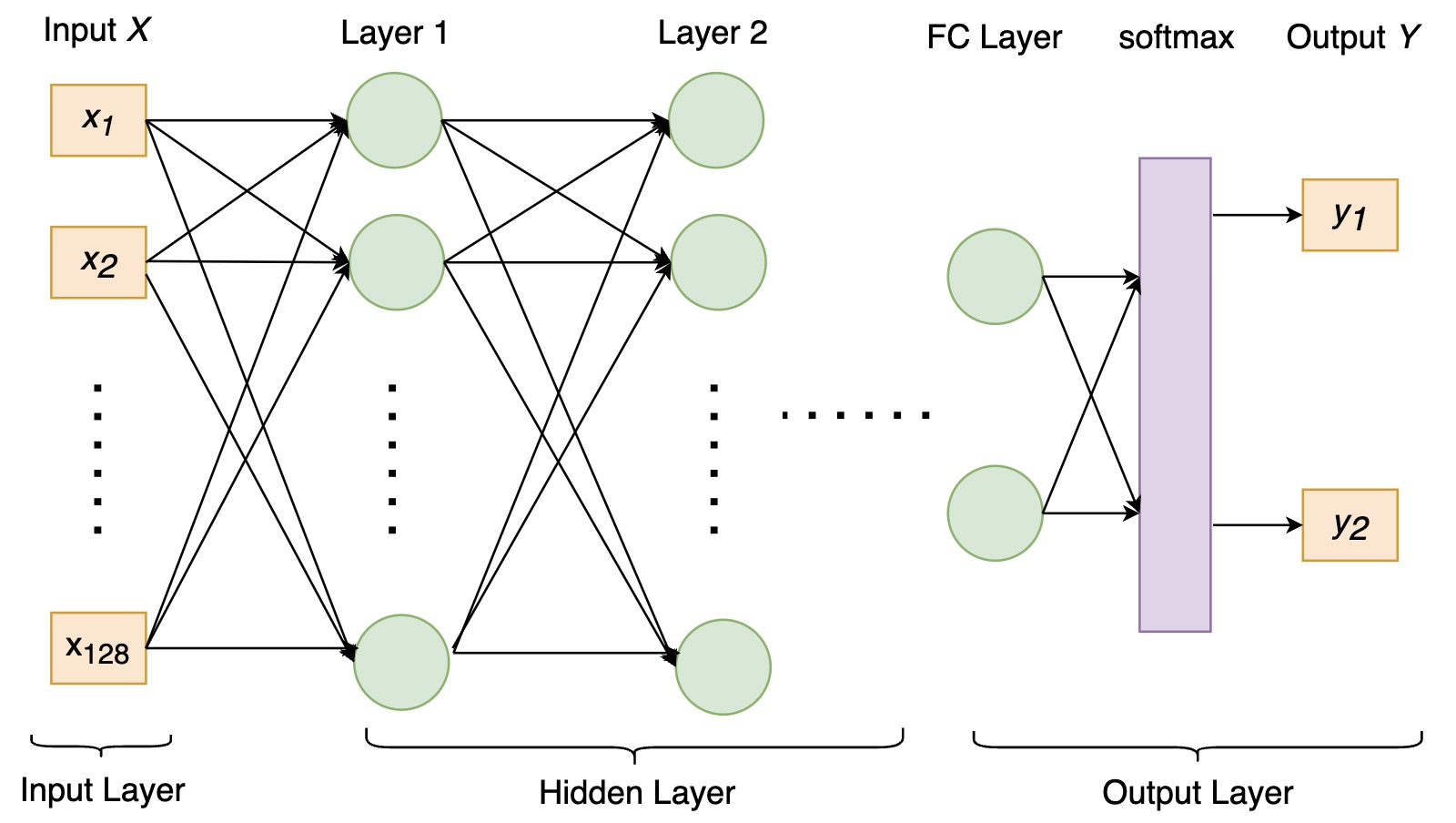}
        \caption{Designed FNN model for NSL-KDD}
        \label{fig:4_28}
    \end{minipage}
    \begin{minipage}[b]{0.3\linewidth}
        \centering
        \includegraphics[width=1\linewidth]{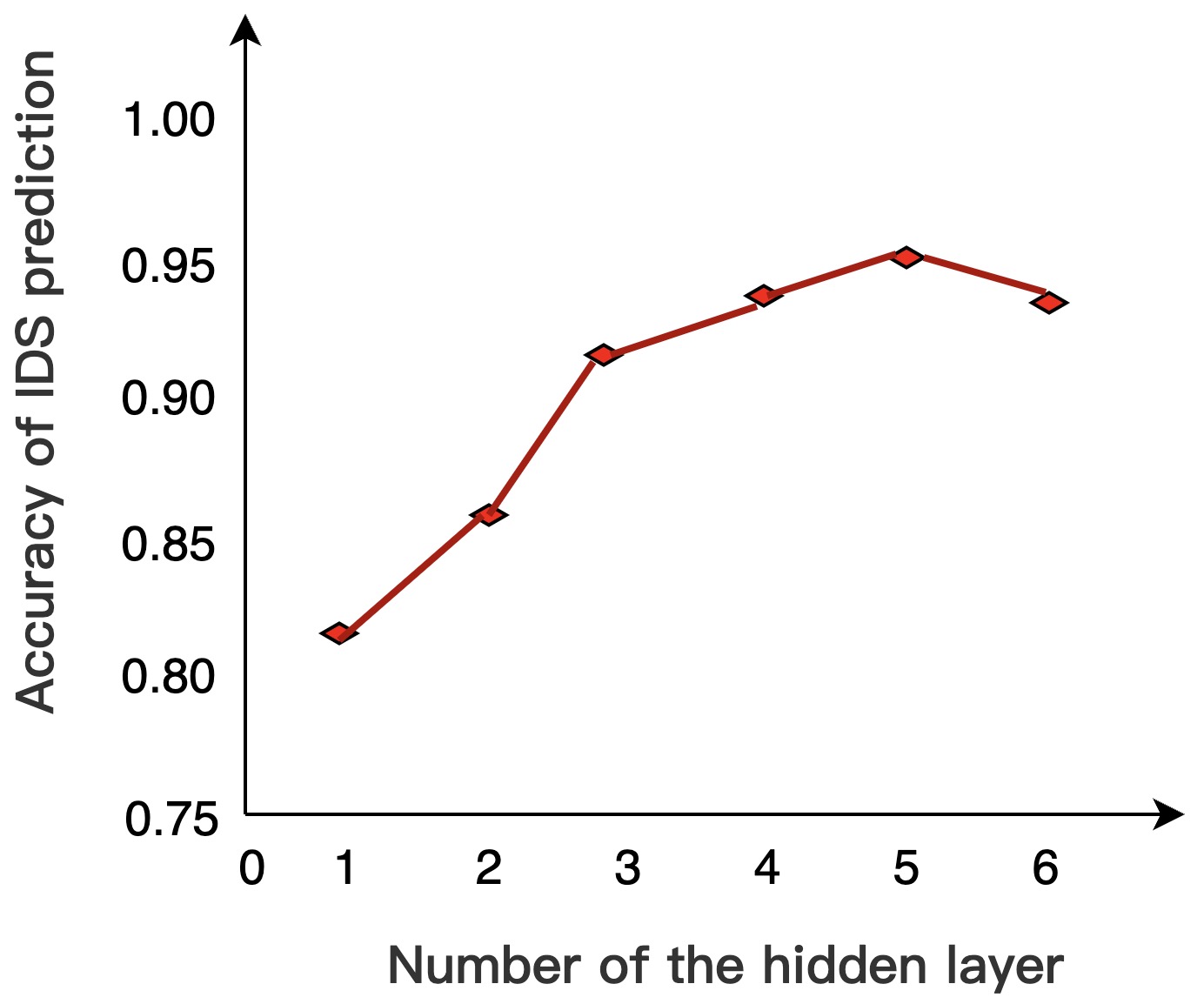}
        \caption{Network layer number and detection accuracy of the proposed FNN model}
        \label{fig:4_29}
    \end{minipage}
    \vspace{-0.5em} 
\end{figure}

\subsection{Adversarial examples detector}
Previous research has found that the adversarial space is a subspace of the original data and is characterized by being a low-probability region within the distribution of the original data. The boundaries of the adversarial space are close to the clean data set in the adversarial direction\cite{b41}. However, they often belong to different data distributions on their submanifolds. Furthermore, it has been further discovered that the transferability of these subspaces to other models increases with the higher number of orthogonal adversarial directions\cite{b27}. The goal of this study is to develop an AE detector for detecting whether the input traffic to IDS is an AE. Due to the low dimensionality of the data and the need to carefully control the magnitude of adversarial perturbations applied to network traffic, the spatial distance characteristics of AEs become challenging to capture. The kernel density estimation (KDE) fails to reflect the spatial characteristics of AEs\cite{b28}, and the performance of the KDE detector is not satisfactory. LID is a method used as an alternative to KDE measurement to characterize the spatial features of examples. It relies on the distribution of internal distances between examples to capture the intrinsic dimensionality of the data. This method has been successfully applied in various domains, such as dimension reduction, similarity search, and anomaly detection. Moreover, advancements have been made in utilizing this method for extracting features of AEs in the field of object detection\cite{b23}. Therefore, it can be considered to create an AE detector by extracting the LID of clean examples and AEs from the traffic and using the statistical differences between them.

Given a data example $x \in X$, let $r$ represent the distance from x to other data examples. The cumulative distribution function of r is denoted as $F(r)$. The LID at r can be defined as:

\begin{equation}
       \mathrm{LID}_{F}(r) \triangleq \lim _{\epsilon \rightarrow 0} \frac{\ln (F((1+\epsilon) \cdot r) / F(r))}{\ln (1+\epsilon)}=\frac{r \cdot F^{\prime}(r)}{F(r)} 
\end{equation}

LID can describe the relative rate at which the cumulative distance function $F(r)$ increases with distance r. In the case of a uniformly distributed sample around $x$, $LID_{F}$ is highly correlated with the dimensionality of the submanifold, and generally, the value of $LID_{F}$ is approximately equal to the dimensionality of the submanifold. To estimate the values of LID, we take $S$ as a small batch of clean examples X, where $r_{ix}$ represents the distance between sample $x$ and the $i-th$ nearest sample in $S$. We then fit a power law to the resulting distribution of distances, resulting in various estimation methods developed for LID, among which the maximum likelihood estimator (MLE) has shown the best performance\cite{b48}. The LID can be represented as:

\begin{equation}
        \widehat{\mathbf{L I D}}(x)=-\left(\frac{1}{k} \sum_{i=1}^{k} \log \frac{r_{i}(x)}{r_{\max }(x)}\right)^{-1}
\end{equation}

The parameter $k$ refers to a clamping factor, which represents the number of nearest neighbors that need to be considered for sample $x$ during computation. Different values of $k$ will result in different LID calculations\cite{b29}, which can impact the performance of the detector. After conducting multiple experiments, we select $k=10$ as the parameter for the detector. Since computing the nearest neighborhood of the entire dataset can be computationally expensive, we select the activation values of the intermediate layers in the neural network to measure the distance between the nth activation layers. Then, LID in our scheme can be regarded as $\left \{LIDd_{1}, LID_{2} ... LID_{n}\right \}$. Finally, we obtain a large number of LID values for two classes of samples and needed to train a classifier that can fit well using the LID of AEs and clean data. In this study, we find that the logistic regression classifier used in the image domain did not perform well in this case. Therefore, in order to improve the classification performance, we experiment with various methods to train the final classifier and find that SVM achieved the highest classification accuracy and exhibited good transferability for attack detection. We observe that the LID of AEs is generally higher than that of clean examples, indicating that AEs reside in a higher-dimensional space with similar spatial characteristics\cite{b30}. By effectively utilizing this similarity and leveraging limited prior knowledge, it is possible to successfully detect the majority of AEs.

In previous research, Anish Athalya et al.\cite{b31} mentioned that the computation of LID involves the calculation of k-nearest neighbors, where minimizing the gradient of k-nearest neighbors distances does not represent the true gradient descent direction in its neighborhood. Therefore, it fails to withstand carefully crafted CW attacks. \kibitz{black}{We infer} that the main reason for this evasion issue is the excessive depth of the deep learning models used for object detection, along with multiple convolutional pooling operations that diminish the linear relationship between neighboring layers. Our DL-based IDS model utilizes fully connected layers with fewer neurons per activation layer, resulting in stronger linear relationships. This enhances the representativeness of distance values calculated by k-nearest neighbors, making the performance of the LID-based AE detector acceptable even against CW attacks.

\begin{center}
\begin{minipage}{0.7\linewidth}

\begin{algorithm}[H]
	\caption{Training phase for LID-based adversarial detector}
	\label{alg:algorithm1}
	\KwIn{ 
		\\$X$: normal examples in dataset;\par 
		$D(x)$: a pre-trained DNN with L hidden layers;\par 
		$k$: the number of nearest neighbors for LID estimation
	}
	\KwOut{
		\\Detector(LID) \tcp{a result from detector}
	}  
	\BlankLine

	$LID_{neg}=[]$ , $LID_{pos}=[]$ \tcp{positive represents AE}
	
	\ForEach{$P_{clean}$ in $X$}{
		\tcp{a minibatch of examples in $X$}
		$P_{adv}$ = adversarial attack $P_{clean}$\;
		$N = \left |  P_{clean} \right |$ \tcp{number of examples}
		$LID_{neg}$, $LID_{pos}$ = zeros[$N,L$]\;
		
		\ForEach{$i$ in $\left [ 1,L\right ]$}{
			$A_{clean}$ = $D^{i}(P_{clean})$\;
			$A_{adv}$ = $D^{i}(P_{adv})$\;
			
			\ForEach{$j \in [1,N]$}{
				$LID_{neg}[j,i]$ = $-\left(\frac{1}{k} \sum_{i=1}^{k} \log \frac{r_{i}(A_{\text {clean}}[j], A_{\text {clean}})}{r_{k}(A_{\text {clean}}[j], A_{\text {clean}})}\right)^{-1}$\;
				$LID_{adv}[j,i]$ = $-\left(\frac{1}{k} \sum_{i=1}^{k} \log \frac{r_{i}(A_{\text {adv}}[j], A_{\text {clean}})}{r_{k}(A_{\text {adv}}[j], A_{\text {clean}})}\right)^{-1}$\;
			}
		}
		$LID_{neg}$.append($LID_{neg}$)\;
		$LID_{pos}$.append($LID_{adv}$)\;
	}
	\Return Detector(LID) = train a SVM classifier on $(LID_{neg}, LID_{pos})$
\end{algorithm}

\end{minipage}
\end{center}
\subsection{ML-based IDS}

\begin{figure}[htbp]
  \centering
  \includegraphics[width=0.5\textwidth,height=6cm]{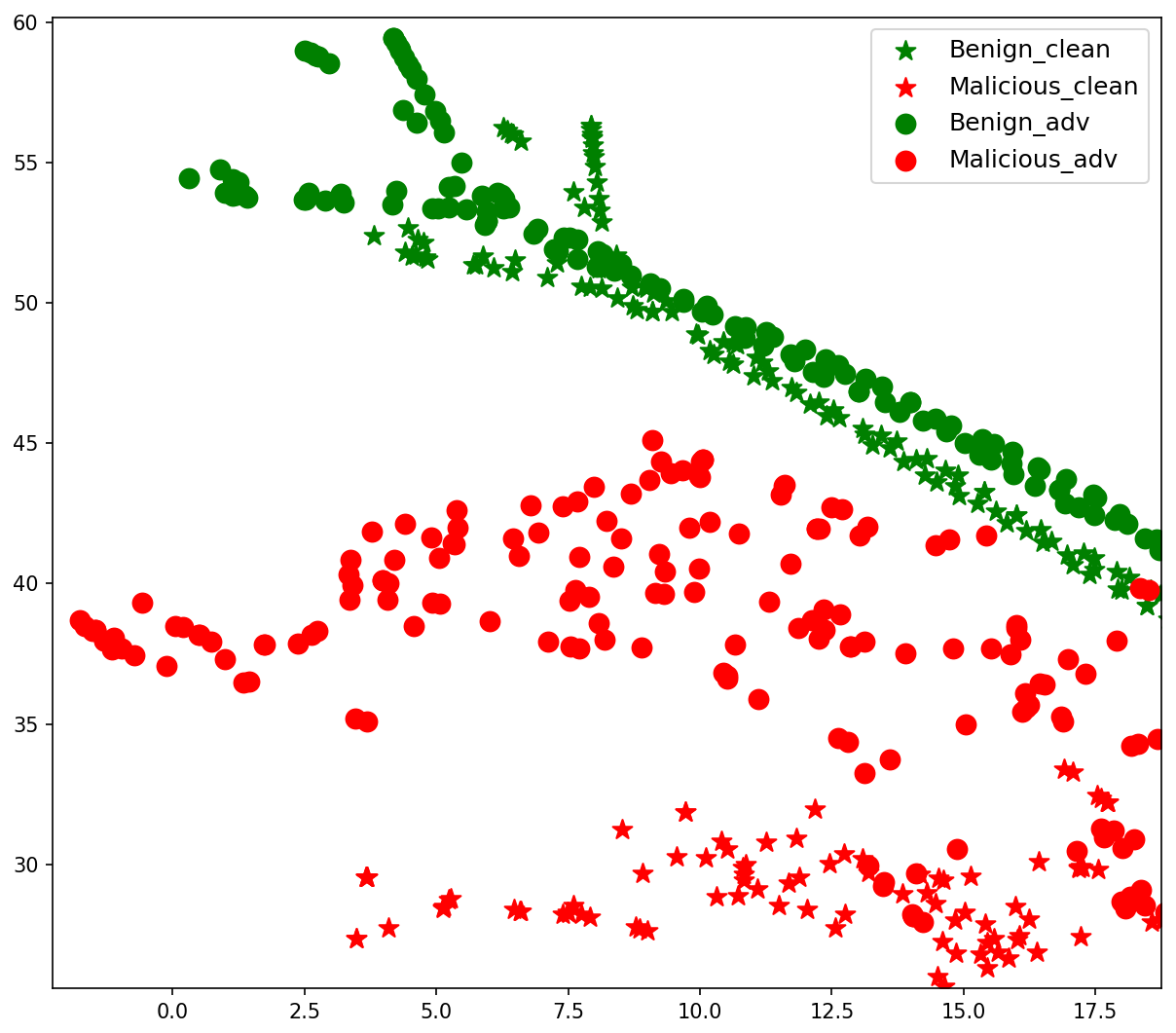}
  \caption{Projection plot of some clean examples of NSL-KDD and  AEs which generated under 10\% perturbation restriction with BIM attack.}
  \label{fig7}
\end{figure}

AEs are generated by adding small perturbations to clean examples. These examples can cross decision boundaries, causing DL models to make incorrect decisions. For example, an AE of malicious traffic may appear benign to the DL-based IDS. However, we have a basis for assuming that the generated AEs are closer in terms of spatial distribution to the original clean class. To validate this hypothesis, we project a subset of the AEs obtained after the attacks onto a two-dimensional plane using \kibitz{black}{t-distributed stochastic neighbor embedding (t-SNE)}, as shown in Figure \ref{fig7}. The figure clearly indicates that the distribution of AEs of malicious traffic is relatively close to that of clean examples of malicious traffic. They reside in the same malicious manifold space, which indirectly supports the notion that they still belong to the same class of traffic. We consider that some machine learning algorithms in the field have the ability to extract and preserve spatial features of samples, and relevant studies have shown that machine learning models exhibit a certain level of robustness and transferability to AEs\cite{b32}. Thus, we attempt to identify an ML model that can still make relatively accurate predictions for AEs. In our experiments, we compare multiple machine learning methods and consider their performance. To comprehensively capture the spatial manifolds of both malicious and benign traffic, we ultimately select the LS algorithm proposed by Zhu et al.\cite{b33} This learning method is a semi-supervised algorithm that transforms samples into a graph representation and utilizes a combination of labeled and unlabeled samples to reveal the intrinsic structure of the sample manifold. By leveraging the relationships between graph nodes, labels are propagated from labeled samples to unlabeled samples. We attempt to analyze the robustness of this algorithm to AEs from the perspective of algorithm principles. The algorithm flow is represented as follows:

\begin{itemize}
\item[1)] 
The algorithm calculates an affinity matrix $W$, which is an $n\ast n$ symmetric matrix in our case. Each element $w_{ij}$ in the matrix represents the estimated distance between $x_{i}$ and $x_{j}$ using a Gaussian kernel, $w_{ij} = \exp \left(-\left\|x_{i}-x_{j}\right\|^{2} / 2 \sigma^{2}\right)$ and $w_{ii}=0$ $(i < n)$.
\item[2)] 
The degree matrix $D$ is computed, which is a diagonal matrix where each element $d_{ii}$ is the sum of $w_{ij}$ for the $i-th$ sample, representing the sum of distances from the $i-th$ sample to all other samples. The elements $d_{ij} = 0$  $(i \ne  j)$.
\item[3)]
Construct a normalized Laplacian matrix: $L =  D^{-\frac{1}{2}}WD^{-\frac{1}{2}}$.
\item[4)]
Iteratively compute $\tilde{Y}_{(t+1)} = \alpha L\tilde{Y}_{(t)} + (1-\alpha)\tilde{Y}_{(0)}$ until it converges, where $\tilde{Y}_{0}$ is an $n\ast c$ matrix and represents the initial labels of the samples, c represents the number of classes to be assigned. $\alpha$ is a clamping factor, and if $\alpha = 0$, the algorithm will always reset the labels to their initial values.
\end{itemize}

The convergence of this iterative algorithm has been proven\cite{b34}. We primarily focus on the robustness of the algorithm when dealing with AEs. By observing the iterative equation mentioned above, it is obvious that it resembles a diffusion problem in a discrete setting. Computing the cost function $\triangle \tilde{Y} = \tilde{Y}_{(t+1)} - \tilde{Y}_{(t)}$, we observe that it is directly proportional to the Laplacian matrix $L$. It can be demonstrated that this computation is equivalent to minimizing the following cost function\cite{b35}:

\begin{equation}
        L(\tilde{Y})=\lVert \tilde{Y}_{L}-Y_{L} \rVert ^2 +\lVert \tilde{Y}_{U}\rVert^2 +\mu(D^{-\frac{1}{2}})^T(D-W)(D^{-\frac{1}{2}}\tilde{Y})
\end{equation}
The first term on the right-hand side of the equation determines the consistency between the predicted labels and the original labels. The second term utilizes a normalization factor to ensure that unlabeled points have a contribution of zero. The third term is employed to enforce smoothness and geometric coherence along the boundaries of the manifold. The variable $\mu$ in this equation is directly proportional to $\alpha$. By minimizing the third term, rapid changes in the labels of high-density region samples will be penalized. The iterative process in the algorithm not only ensures convergence but also imposes constraints on the labels of samples in dense regions. This is likely to contribute to the robustness of the model when facing AEs.

We have observed that both the Label-propagation and LS algorithms have similar procedures. However, in terms of performance in predicting AEs, LS demonstrates superior performance. From this observation, we can speculate that the robustness of LS may stem from the normalization Laplacian matrix in the third step. In the experiments conducted in this study, the influence of different distances between samples was taken into consideration during the iteration process of the $\tilde{Y}_{(t)}$ matrix in the LS algorithm. It was assumed that if there are multiple samples with determined labels in the immediate vicinity of sample $x_{1}$, the label of sample $x_{1}$ would be determined by the labels of the surrounding samples during the iteration. The inter-sample distance serves as a crucial factor in determining the contribution of labels, with closer samples exerting a more significant influence on label computation. Consequently, even in the presence of perturbations, the label of the AE ${x}'$  will continue to be influenced by neighboring samples. The algorithm will exhibit robustness if the perturbations applied to the AE remain within a sufficiently small range. By utilizing the Rayleigh quotient\cite{b36}, $R(A,x) = \frac{x^{T}Ax}{x^{T}x}$, the upper and lower bounds of the eigenvalues of the normalized Laplacian matrix can be obtained.

\begin{align}
        R(L_{sym}, g) &= \frac{g^T D^{-\frac{1}{2}} L D^{-\frac{1}{2}} g}{g^T g} 
= \frac{f^T L f}{\left(D^{\frac{1}{2}}f\right)^T \left(D^{\frac{1}{2}}f\right)} 
= \frac{\sum_v f^2(v) d_v - \sum_{v} \sum_{v-u} f(u)f(v)}{\sum_v f^2(v) d_v} 
= \frac{\sum_{v-u}(f(u)-f(v))^2}{\sum_v f^2(v) d_v}
\end{align}
From the result, it can be observed that the numerator is the sum of the Euclidean distances of vector $f$. By penalizing this numerator, a smoothing effect can be achieved. Related research provided theoretical evidence for the existence of a strong relation between large robustness and small curvature\cite{b37}.

\section{Experiments}
\subsection{Dataset}
Since 1999, KDD’99 has been the most widely used data set for the evaluation of anomaly detection methods. However, the most important deficiency of the KDD dataset is the huge number of redundant records. In 2010, Mahbod Tavallaee proposed a new data set NSL-KDD\cite{b38}, which solved the problems caused by KDD. As a result, we choose NSL-KDD as our dataset to test the DLL-IDS system we proposed in this paper. This training dataset consists of approximately 4,000,000 single connection vectors each of which contains 41 features and is labeled as either normal or an attack. The attacks fall into one of the following four categories: Denial of Service Attack (DoS), User to Root Attack (U2R), Remote to Local Attack (R2L), and Probing Attack of which each contains more attack sub-categories. In this study, IDS considered traffic as a binary classification problem. We select 120,000 training samples and 25,000 testing samples from the original data. Due to space limitations, in the subsequent presentation of experimental results, we choose representative DOS attack data to showcase the effectiveness of our IDS. The defense performance against the other three types of attacks is also similar.

\kibitz{black}{In order to ensure the reliability of our experimental conclusions, we also choose an additional dataset, CICIDS2018, which was published by Canadian Institute for Cybersecurity (CIC) for evaluation. This dataset resembles real-world Packet Capture data, it contains normal network traffic and several classes of intrusions, including Brute-force, Heartbleed, DDoS, Web attacks, and so on. We preprocess the data by removing four features with a high number of infinite values and missing values. This resulted in retaining 78 features for training. For this dataset, we opt for a deep learning model structure using Convolutional Neural Networks (CNN) to train a DL-based IDS. We use 1.5 million network flows as the training dataset and 300,000 network flows as the testing dataset. Similar to what was mentioned above with NSL-KDD, we continue to focus on the DDOS attack type when presenting our experimental results.}

\subsection{AE Generation}

In the initial NSL-KDD, there are a total of 41 feature dimensions, which we divide into persistent features and dynamic features based on whether the feature vector changes for some reason in the real world. We select six persistent features, including protocol\_type, service, flag, land, and 35 dynamic features.
In order to incorporate persistent features into the model training, we choose to normalize them and append them to the end of each sample, expanding the original 41-dimensional feature vector to a 128-dimensional feature vector. When calculating the impact of perturbation on the original data, we only add perturbations to the 35 dynamic features. We hope this method respects the nature of regular tabular data, encompassing both discrete and continuous features\cite{b49}. 
\kibitz{black}{For the data from CICIDS2018, after extracting relevant network traffic information using the official CICFlowMeter tool, we remove certain feature dimensions that have missing values and infinite values. We reshape it into a square feature matrix with dimensions of $9 \times 9$ by zero padding. Subsequently, we employ a CNN for model training and conduct AE attack testing.}

To better align with real-world scenarios\cite{b39}, the magnitude of each perturbation is constrained to be within 20\% of the original data. The following AEs are generated by attacking DL models using different methods while controlling the restriction of perturbations. Additionally, adversarial attacks are only applied to examples that can be correctly predicted. We evaluate the transferability of AEs between different models by statistically analyzing the performance of other models on these AEs. We assess the detection capability of the LID-based AE detector and the overall robustness of the DLL-IDS system through this analysis.
\subsubsection{AE Generation of CICIDS2018}

\subsection{Evaluation}
\begin{table}[htbp]
\caption{The DL-based IDS Accuracy under adversarial attacks}
\label{The DL-based IDS Accuracy under adversarial attacks}
\centering
\begin{tabular}{|c|c|c|c|c|c|}
\hline
Dataset &
  \begin{tabular}[c]{@{}c@{}}Perturbation\\ restriction(\%)\end{tabular} &
  \begin{tabular}[c]{@{}c@{}}FGSM\\ Acc(\%)\end{tabular} &
  \begin{tabular}[c]{@{}c@{}}BIM\\ Acc(\%)\end{tabular} &
  \begin{tabular}[c]{@{}c@{}}DEEPFOOL\\ Acc(\%)\end{tabular} &
  \begin{tabular}[c]{@{}c@{}}CW\\ Acc(\%)\end{tabular} \\ \hline
\multicolumn{1}{|c|}{\multirow{6}{*}{NSL-KDD}} &
  \multicolumn{1}{c|}{0} &
  \multicolumn{1}{c|}{94.3} &
  \multicolumn{1}{c|}{94.3} &
  \multicolumn{1}{c|}{94.3} &
  \multicolumn{1}{c|}{94.3} \\ \cline{2-6} 
\multicolumn{1}{|c|}{} &
  \multicolumn{1}{c|}{2.0} &
  \multicolumn{1}{c|}{83.6} &
  \multicolumn{1}{c|}{82.1} &
  \multicolumn{1}{c|}{73.0} &
  \multicolumn{1}{c|}{63.4} \\ \cline{2-6} 
\multicolumn{1}{|c|}{} &
  \multicolumn{1}{c|}{5.0} &
  \multicolumn{1}{c|}{78.3} &
  \multicolumn{1}{c|}{71.5} &
  \multicolumn{1}{c|}{56.7} &
  \multicolumn{1}{c|}{52.2} \\ \cline{2-6} 
\multicolumn{1}{|c|}{} &
  \multicolumn{1}{c|}{10.0} &
  \multicolumn{1}{c|}{74.5} &
  \multicolumn{1}{c|}{62.3} &
  \multicolumn{1}{c|}{31.4} &
  \multicolumn{1}{c|}{18.9} \\ \cline{2-6} 
\multicolumn{1}{|c|}{} &
  \multicolumn{1}{c|}{20.0} &
  \multicolumn{1}{c|}{65.2} &
  \multicolumn{1}{c|}{23.8} &
  \multicolumn{1}{c|}{15.1} &
  \multicolumn{1}{c|}{5.7} \\ \cline{2-6} 
\multicolumn{1}{|c|}{} &
  \multicolumn{1}{c|}{None} &
  \multicolumn{1}{c|}{46.9} &
  \multicolumn{1}{c|}{7.2} &
  \multicolumn{1}{c|}{0} &
  \multicolumn{1}{c|}{0} \\ \hline
\multicolumn{1}{|c|}{\multirow{6}{*}{CICIIDS2018}} &
  \multicolumn{1}{c|}{0} &
  \multicolumn{1}{c|}{99.3} &
  \multicolumn{1}{c|}{99.3} &
  \multicolumn{1}{c|}{99.3} &
  \multicolumn{1}{c|}{99.3} \\ \cline{2-6} 
\multicolumn{1}{|c|}{} &
  \multicolumn{1}{c|}{2.0} &
  \multicolumn{1}{c|}{79.8} &
  \multicolumn{1}{c|}{78.7} &
  \multicolumn{1}{c|}{69.1} &
  \multicolumn{1}{c|}{50.2} \\ \cline{2-6} 
\multicolumn{1}{|c|}{} &
  \multicolumn{1}{c|}{5.0} &
  \multicolumn{1}{c|}{57.0} &
  \multicolumn{1}{c|}{53.4} &
  \multicolumn{1}{c|}{52.3} &
  \multicolumn{1}{c|}{40.0} \\ \cline{2-6} 
\multicolumn{1}{|c|}{} &
  \multicolumn{1}{c|}{10.0} &
  \multicolumn{1}{c|}{28.6} &
  \multicolumn{1}{c|}{23.3} &
  \multicolumn{1}{c|}{17.9} &
  \multicolumn{1}{c|}{17.3} \\ \cline{2-6} 
\multicolumn{1}{|c|}{} &
  \multicolumn{1}{c|}{20.0} &
  \multicolumn{1}{c|}{6.1} &
  \multicolumn{1}{c|}{2.0} &
  \multicolumn{1}{c|}{0} &
  \multicolumn{1}{c|}{0} \\ \cline{2-6} 
\multicolumn{1}{|c|}{} &
  \multicolumn{1}{c|}{None} &
  \multicolumn{1}{c|}{0} &
  \multicolumn{1}{c|}{0} &
  \multicolumn{1}{c|}{0} &
  \multicolumn{1}{c|}{0} \\ \hline
\end{tabular}
\end{table}
As shown in Table \ref{The DL-based IDS Accuracy under adversarial attacks}, the DL-based IDS's classification accuracy is demonstrated when confronted with AEs generated using different attack methods under varying perturbation constraints. Since all clean examples were correctly classified, we define the prediction accuracy on attacked dataset as \textit{Acc}. Based on past experience, it is known that the attack intensity of these four attack methods increases from left to right\cite{b42}. It is evident that as the attack intensity increases, the prediction accuracy of the model decreases for higher attack intensities under the same perturbation constraint. For instance, under the same perturbation constraint of 5\%, the DL model of NSL-KDD achieves a prediction accuracy of 78.3\% for AEs generated by the FGSM algorithm. However, the CW attack, which has been found to be the most effective attack in the current investigation, results in a prediction success rate of only 52.2\% for the DL model of NSL-KDD under the same perturbation constraint. Similar to this scenario, when subjected to the same type of attack, there is an inverse relationship between the perturbation constraint and the prediction success rate. AEs generated under looser perturbation constraints also lead to a decrease in the model's prediction accuracy. From the \kibitz{black}{part of CICIDS2018}, it becomes evident that the prediction success rate of the FGSM attack decreases from 79.8\% to 6.1\% under different perturbation constraints. Furthermore, the last row indicates that if there are no limitations on the perturbation, it becomes relatively easy to achieve a near-zero prediction success rate, resulting in nearly perfect attack success. However, the rationality of such adversarial examples generated without any perturbation constraints warrants further consideration.


Now, we have generated a large number of AEs that perform well on the attacking side using different attack methods. We need to evaluate the effectiveness of the LID AE detector designed in the previous discussion. In the final selection of the fitting algorithm for the AE detector based on LID values, we choose logistic regression (LGR), decision tree classifier (DTC), bernoulli naive bayes
(BNB), and support vector machine (SVM) as the four algorithms. Their performance \kibitz{black}{on NSL-KDD} is shown in Table \ref{The detection accuracy of the LID-based AE detector when using different fitting algorithms.}. It can be observed that the BNB algorithm is almost ineffective, with a very high False Positive Rate (FPR). Among the remaining three algorithms, SVM and DCT exhibit significantly better performance than LGR. Moreover, SVM and DCT show similar performance. Therefore, we ultimately select SVM as the fitting algorithm due to its relatively superior performance. To ensure the informative nature of the experimental results, this study selects DB \cite{b15} as an artifact, a method that has shown relatively good performance in IDS AE detection, for comparison. Some researchers consider DB as one of the best solutions for detecting AEs in network traffic\cite{b2}. The principle of DB is to explore whether an input $x$ is near the decision boundary of the IDS model in order to detect AEs. More specifically, by adding small Gaussian noise to $x$ and observing the corresponding changes in $y$, it can be determined if $x$ is highly likely to be an AE. Table \ref{Performance comparison of AE detectors} shows the performances of our LID Detector and DB Detector in AEs generated from the DL-based IDS model by different attacks. We compare the detection accuracies of the two detection methods using two types of samples: AEs and clean examples. From the perspective of AEs, we generate them with perturbation sizes set at 5\%, 10\%, and 20\% for each type of attack. 
\begin{table}[htbp]
\caption{The detection accuracy of the LID-based AE detector when using different fitting algorithms.}
\label{The detection accuracy of the LID-based AE detector when using different fitting algorithms.}
\centering
\resizebox{\linewidth}{!}{
\begin{tabular}{|c|cccc|cccc|cccc|cccc|cccc|}
\hline
\begin{tabular}[c]{@{}c@{}}Attack\\ Method\end{tabular} & \multicolumn{4}{c|}{CLEAN}                                                              & \multicolumn{4}{c|}{FGSM}                                                                & \multicolumn{4}{c|}{BIM}                                                                 & \multicolumn{4}{c|}{DEEPFOOL}                                                            & \multicolumn{4}{c|}{CW}                                                                  \\ \hline
\begin{tabular}[c]{@{}c@{}}Method\\ in LID\end{tabular} & \multicolumn{1}{c|}{LGR}  & \multicolumn{1}{c|}{DCT}  & \multicolumn{1}{c|}{BNB} & SVM  & \multicolumn{1}{c|}{LGR}  & \multicolumn{1}{c|}{DCT}  & \multicolumn{1}{c|}{BNB}  & SVM  & \multicolumn{1}{c|}{LGR}  & \multicolumn{1}{c|}{DCT}  & \multicolumn{1}{c|}{BNB}  & SVM  & \multicolumn{1}{c|}{LGR}  & \multicolumn{1}{c|}{DCT}  & \multicolumn{1}{c|}{BNB}  & SVM  & \multicolumn{1}{c|}{LGR}   & \multicolumn{1}{c|}{DCT}  & \multicolumn{1}{c|}{BNB} & SVM  \\ \hline
ACC(\%)                                                 & \multicolumn{1}{c|}{85.7} & \multicolumn{1}{c|}{90.2} & \multicolumn{1}{c|}{5.1} & 89.0 & \multicolumn{1}{c|}{68.6} & \multicolumn{1}{c|}{88.4} & \multicolumn{1}{c|}{99.6} & 87.9 & \multicolumn{1}{c|}{70.1} & \multicolumn{1}{c|}{85.4} & \multicolumn{1}{c|}{99.9} & 89.8 & \multicolumn{1}{c|}{81.8} & \multicolumn{1}{c|}{92.1} & \multicolumn{1}{c|}{98.1} & 95.1 & \multicolumn{1}{c|}{72.16} & \multicolumn{1}{c|}{86.1} & \multicolumn{1}{c|}{100} & 85.5 \\ \hline
\end{tabular}}
\end{table}

\begin{table}[htbp]
\caption{Performance comparison of AE detectors}
\label{Performance comparison of AE detectors}
\centering
\resizebox{\linewidth}{!}{
\begin{tabular}{|c|l|c|ccc|ccc|ccc|ccc|}
\hline
\multirow{2}{*}{Dataset} &
  \multicolumn{1}{c|}{\multirow{2}{*}{\begin{tabular}[c]{@{}c@{}}Detection\\ Method\end{tabular}}} &
  CLEAN &
  \multicolumn{3}{c|}{FGSM} &
  \multicolumn{3}{c|}{BIM} &
  \multicolumn{3}{c|}{DEEPFOOL} &
  \multicolumn{3}{c|}{CW} \\ \cline{3-15} 
 &
  \multicolumn{1}{c|}{} &
  0 &
  \multicolumn{1}{c|}{5\%} &
  \multicolumn{1}{c|}{10\%} &
  20\% &
  \multicolumn{1}{c|}{5\%} &
  \multicolumn{1}{c|}{10\%} &
  20\% &
  \multicolumn{1}{c|}{5\%} &
  \multicolumn{1}{c|}{10\%} &
  20\% &
  \multicolumn{1}{c|}{5\%} &
  \multicolumn{1}{c|}{10\%} &
  20\% \\ \hline
\multirow{2}{*}{NSL-KDD} &
  LID &
  96.9 &
  \multicolumn{1}{c|}{83.3} &
  \multicolumn{1}{c|}{88.1} &
  90.1 &
  \multicolumn{1}{c|}{79.1} &
  \multicolumn{1}{c|}{82.9} &
  85.7 &
  \multicolumn{1}{c|}{83.2} &
  \multicolumn{1}{c|}{95.1} &
  96.6 &
  \multicolumn{1}{c|}{72.8} &
  \multicolumn{1}{c|}{78.1} &
  83.4 \\ \cline{2-15} 
 &
  Artifact &
  82.1 &
  \multicolumn{1}{c|}{66.4} &
  \multicolumn{1}{c|}{65.5} &
  66.7 &
  \multicolumn{1}{c|}{77.4} &
  \multicolumn{1}{c|}{85.1} &
  88.5 &
  \multicolumn{1}{c|}{68.9} &
  \multicolumn{1}{c|}{69.8} &
  73.5 &
  \multicolumn{1}{c|}{52.4} &
  \multicolumn{1}{c|}{76.9} &
  89.6 \\ \hline
\multirow{2}{*}{CICIDS2018} &
  LID &
  96.3 &
  \multicolumn{1}{c|}{87.1} &
  \multicolumn{1}{c|}{95.0} &
  74.4 &
  \multicolumn{1}{c|}{94.5} &
  \multicolumn{1}{c|}{87.5} &
  79.0 &
  \multicolumn{1}{c|}{82.6} &
  \multicolumn{1}{c|}{89.0} &
  92.6 &
  \multicolumn{1}{c|}{91.3} &
  \multicolumn{1}{c|}{98.4} &
  88.2 \\ \cline{2-15} 
 &
  Artifact &
  88.3 &
  \multicolumn{1}{c|}{67.5} &
  \multicolumn{1}{c|}{82.7} &
  60.5 &
  \multicolumn{1}{c|}{76.5} &
  \multicolumn{1}{c|}{77.1} &
  60.2 &
  \multicolumn{1}{c|}{63.4} &
  \multicolumn{1}{c|}{75.2} &
  78.7 &
  \multicolumn{1}{c|}{35.5} &
  \multicolumn{1}{c|}{42.3} &
  65.0 \\ \hline
\end{tabular}}
\end{table}

Based on Table \ref{Performance comparison of AE detectors}, it can be observed that the LID method implemented in this study exhibits superior accuracy compared to Artifact in nearly all cases. Our experimental findings suggest that this discrepancy may be attributed to the fact that Artifact necessitates the consideration of various attack method characteristics in order to fine-tune the algorithm's parameters. However, in real-world network environments, it becomes challenging for IDS to anticipate the specific attack method chosen by adversaries when confronted with GB and WB attacks. Achieving real-time parameter adjustment for the model is almost impossible, making it impractical in real-world scenarios. Furthermore, we have observed that for the same type of attack, AEs with larger perturbations are more likely to be detected.

\begin{table}[htbp]
\caption{Transferability of adversarial attacks}
\label{Transferability of adversarial attacks}
\centering
\begin{tabular}{|cc|ccccc|}
\hline
\multicolumn{2}{|c|}{}           & \multicolumn{5}{c|}{Acc(\%)}                                                                                         \\ \hline
\multicolumn{1}{|c|}{Models} &
  \begin{tabular}[c]{@{}c@{}}Acc\\ (\%)\end{tabular} &
  \multicolumn{1}{l|}{FGSM} &
  \multicolumn{1}{l|}{BIM} &
  \multicolumn{1}{l|}{DEEPFOOL} &
  \multicolumn{1}{l|}{CW} &
  \multicolumn{1}{l|}{Overall} \\ \hline
\multicolumn{1}{|c|}{LGR} & 92.5 & \multicolumn{1}{c|}{80.2} & \multicolumn{1}{c|}{69.2} & \multicolumn{1}{c|}{38.8} & \multicolumn{1}{c|}{30.0} & 54.6 \\ \hline
\multicolumn{1}{|c|}{BNB} & 87.2 & \multicolumn{1}{c|}{83.6} & \multicolumn{1}{c|}{76.7} & \multicolumn{1}{c|}{67.0} & \multicolumn{1}{c|}{83.1} & 77.6 \\ \hline
\multicolumn{1}{|c|}{SVM} & 89.4 & \multicolumn{1}{c|}{85.1} & \multicolumn{1}{c|}{61.3} & \multicolumn{1}{c|}{43.3} & \multicolumn{1}{c|}{38.9} & 57.1 \\ \hline
\multicolumn{1}{|c|}{DTC} & 85.0 & \multicolumn{1}{c|}{62.1} & \multicolumn{1}{c|}{57.3} & \multicolumn{1}{c|}{68.2} & \multicolumn{1}{c|}{60.9} & 61.2 \\ \hline
\multicolumn{1}{|c|}{KNN} & 95.7 & \multicolumn{1}{c|}{88.0} & \multicolumn{1}{c|}{82.4} & \multicolumn{1}{c|}{64.4} & \multicolumn{1}{c|}{56.9} & 72.9 \\ \hline
\multicolumn{1}{|c|}{LS}  & 94.3 & \multicolumn{1}{c|}{88.9} & \multicolumn{1}{c|}{82.5} & \multicolumn{1}{c|}{74.7} & \multicolumn{1}{c|}{77.8} & 81.0 \\ \hline
\end{tabular}
\end{table}

It is known that the generated AEs not only cause misclassifications in the targeted DNN model but can also transfer to other models. In order to find an ML-based IDS model with higher robustness, we investigated multiple machine learning models, including KNN, LGR, BNB, DTC, SVM, and LS. We evaluate their performance when faced with AEs generated using DNN models. As shown in Table \ref{Transferability of adversarial attacks}, we present the result on NSL-KDD, LS stands out in terms of robustness compared to other models. After conducting individual tests on each module, it is important to also focus on the strengths of the framework itself, specifically its ability to discriminate malicious samples. To the best of our knowledge, no previous research has addressed how to enhance IDS systems in terms of detection of the inherent maliciousness of traffic AEs. Therefore, in this study, we solely compare the performance of our framework, DLL-IDS, with the baseline model (DL-based IDS) in terms of prediction results. We randomly select 2000 samples for each of the CLEAN, FGSM, BIM, DEEPFOOL, and CW attack methods, with the perturbations controlled within 10\%. The performance of the two IDS models will be compared, we use true positive (TP), false positive (FP), true negative (TN), and false negative (FN) as basic parameters and evaluation metrics of malicious traffic detection are defined as follows:
\begin{itemize}
\item Precision indicates how many correct classifications in all results that are identified as positive: $P = \frac{TP}{TP+FP} $.
\item Recall indicates the proportion of correctly identified malicious traffic over all malicious traffic, formulated as $R = \frac{TP}{TP+FN} $. 
\item F1 score indicates the harmonic mean of precision and recall, formulated as: $F_{1} = \frac{2\cdot P \cdot R}{P+R}$.
\item Accuracy indicates the percentage of correctly classified normal traffic and malicious traffic over all examples: $Acc = \frac{TP+TN}{TP+TN+FP+FN}$.
\end{itemize}

The experimental results are shown in Tables \ref{Performance comparison between DLL-IDS and Baseline on NSL-KDD} and \ref{Performance comparison between DLL-IDS and Baseline on CICIDS2018}. It is demonstrated that the DLL-IDS designed in this study effectively resists various adversarial attacks. The accuracy of AE after CW attack increases from 17.9\% to 71.7\% on NSL-KDD, as shown in Table \ref{Performance comparison between DLL-IDS and Baseline on NSL-KDD}. \kibitz{black}{According to Table \ref{Performance comparison between DLL-IDS and Baseline on CICIDS2018}, it is evident that DLL-IDS on the CICIDS2018 dataset has significantly superior performance compared to the baseline. When confronted with various types of AE attacks, accuracy rates exceed 60\%. This indicated that DLL-IDS demonstrates adaptability to diverse model architectures and datasets of varying specifications.} Furthermore, the overall accuracy of the IDS for clean samples does not significantly decrease. In summary, the DLL-IDS framework greatly improves the performance of DL-based IDS under adversarial attacks, achieving high accuracy with low resource consumption.

\begin{table}[htbp]
\caption{Performance comparison between DLL-IDS and Baseline on NSL-KDD}
\label{Performance comparison between DLL-IDS and Baseline on NSL-KDD}
\centering
\begin{tabular}{|c|c|c|c|c|c|}
\hline
Detection Method          & Test DATA & Precision(\%) & Recall(\%) & F1 score(\%) & Accuracy(\%) \\ \hline
\multirow{5}{*}{DLL-IDS}   & CLEAN     & 87.5          & 99.5       & 93.1         & 91.8         \\ \cline{2-6} 
                          & FGSM      & 89.8          & 95.6       & 92.6         & 90.2         \\ \cline{2-6} 
                          & BIM       & 85.9          & 89.2       & 87.5         & 86.2         \\ \cline{2-6} 
                          & DEEPFOOL  & 69.8          & 98.4       & 81.7         & 72.2         \\ \cline{2-6} 
                          & CW        & 85.7          & 67.3       & 75.3         & 71.7         \\ \hline
\multirow{5}{*}{Baseline} & CLEAN     & 88.0          & 99.3       & 93.3         & 92.1         \\ \cline{2-6} 
                          & FGSM      & 87.2          & 71.8       & 78.7         & 75.7         \\ \cline{2-6} 
                          & BIM       & 84.0          & 59.2       & 69.4         & 67.3         \\ \cline{2-6} 
                          & DEEPFOOL  & 44.8          & 41.0       & 42.8         & 32.3         \\ \cline{2-6} 
                          & CW        & 26.5          & 16.5       & 20.3         & 17.9         \\ \hline
\end{tabular}
\end{table}

\begin{table}[htbp]
\caption{Performance comparison between DLL-IDS and Baseline on CICIDS2018}
\label{Performance comparison between DLL-IDS and Baseline on CICIDS2018}
\centering
\begin{tabular}{|c|c|c|c|c|c|}
\hline
Detection Method          & Test DATA & Precision(\%) & Recall(\%) & F1 score(\%) & Accuracy(\%) \\ \hline
\multirow{5}{*}{DLL-IDS}   & CLEAN     & 100.0          & 96.7       & 98.3         & 98.1         \\ \cline{2-6} 
                          & FGSM      & 97.5          & 94.4       & 95.9         & 95.0        \\ \cline{2-6} 
                          & BIM       & 87.6          & 91.6       & 89.5         & 88.5         \\ \cline{2-6} 
                          & DEEPFOOL  & 54.0          & 92.9       & 68.3         & 71.5         \\ \cline{2-6} 
                          & CW        & 59.4          & 73.5       & 65.7         & 63.7         \\ \hline
\multirow{5}{*}{Baseline} & CLEAN     & 100.0         & 97.4       & 98.6         & 98.5         \\ \cline{2-6} 
                          & FGSM      & 25.3         & 40.2       & 31.1         & 32.0         \\ \cline{2-6} 
                          & BIM       & 24.7          & 28.5       & 26.5         & 22.5         \\ \cline{2-6} 
                          & DEEPFOOL  & 20.3          & 30.6       & 24.4         & 18.5         \\ \cline{2-6} 
                          & CW        & 12.3          & 17.2       & 14.3         & 17.1         \\ \hline
\end{tabular}
\end{table}
\section{Conclusion}

The present study focuses on the classification requirements of real-world traffic samples that serve as AEs, starting from the AE defense of IDS. Four representative adversarial attack algorithms were selected for adversarial attacks, imposing strict constraints on the features of network traffic. The success rates of adversarial example attacks were investigated under different perturbation conditions. After discovering the highly destructive nature of the attacks, we proceed to design a novel DLL-IDS system framework to defend against such attacks, consisting of three components: DL-based IDS, AE detector, and ML-based IDS. During the study of AEs in traffic, we observe that they exhibited inconsistent spatial distribution characteristics compared to clean examples. As a result, we introduce the LID  method to construct an AE detector. This detector achieved a significantly high detection rate and required only a small amount of prior knowledge to successfully detect the majority of AEs. Meanwhile, we exploit the low attack transferability between ML models and traditional DL models to find a robust ML model that assists us in determining the maliciousness of AEs. In our experiments, we observe a significant improvement in the performance of the IDS when subject to adversarial attacks after implementing the DLL-IDS framework. The accuracy, as measured by the highest successful attack rate using the CW attack, increased from 17.9\% to 71.7\%. The framework proposed in this paper has broad applicability to any scenario where attention is required on both the intrinsic properties of the samples and their adversarial nature. Theoretically, this framework can be adapted to any form of data in such scenarios. We hope that future research will increasingly focus on the security issues introduced by deep learning models in other domains.

\section{ACKNOWLEDGMENTS}
This research is supported in part by the National Key Research and Development Program of China, "Dynamically Scalable Mimic Computing Systems and Construction Methods" (Project No. 2022YFB4500900).




\bibliographystyle{unsrt}



\end{document}